\documentclass[]{interact}

\usepackage{epstopdf}
\usepackage{subfigure}
\usepackage{multirow}
\usepackage{float}
\usepackage{blindtext}
\usepackage[utf8]{inputenc}
\usepackage[numbers,sort&compress]{natbib}
\bibpunct[, ]{[}{]}{,}{n}{,}{,}
\newcommand{\f}{\operatorname}

\theoremstyle{plain}

\theoremstyle{definition}

\theoremstyle{remark}

\begin{document}

\articletype{ }

\title{Reliability-centered maintenance: analyzing failure in harvest sugarcane machine using some generalizations of the Weibull distribution}

\author{Pedro Luiz Ramos$^{\rm a}$$^{\ast}$\thanks{$^\ast$Corresponding author. Email: pedrolramos@usp.br
\vspace{6pt}}, Diego Nascimento$^{\rm a}$, Camila Cocolo$^{\rm a}$,  Márcio José Nicola$^{\rm a}$ \\ Carlos Alonso$^{\rm a}$,  Luiz Gustavo Ribeiro, André Ennes$^{\rm a}$  and Francisco Louzada$^{\rm a}$\\ \vspace{6pt}  $^{a}${Institute of Mathematical Science and Computing, University of S\~ao Paulo, S\~ao Carlos, Brazil} }

\maketitle

\begin{abstract}

In this study we considered five generalizations of the standard Weibull distribution to describe the lifetime of two important components of harvest sugarcane machines. 
The harvesters considered in the analysis does the harvest of an average of 20 tons of sugarcane per hour and their malfunction may lead to major losses, therefore, an effective maintenance approach is of main interesting for cost savings. 
For the considered distributions, mathematical background is presented. Maximum likelihood is used for parameter estimation. Further, different discrimination procedures were used to obtain the best fit for each component. At the end, we propose a maintenance scheduling for the components of the harvesters using predictive analysis.
\end{abstract}


\section{Introduction}

The arrival of the sugarcane culture in Brazil has had a significant impact on the national economy, which led the country to become the largest producer in the world \cite{Brazil2017}. Its sub-products are used in the food and chemical industries, as well as in electricity generation and fuel production. Mechanized harvesting is one of the most important stages in the sugar and ethanol mills since it must provide the raw material with quality, time and competitive costs for later processing. Among the used machines in the mechanized harvest, the harvesters stand out for having a large number of corrective stops, given the functionality in such extreme environmental conditions. In addition, its operation is in a regime of 24 hours on the workdays, impacting on fatigue and wear of their parts. During operation, the harvester removes an average of 20 tons of sugarcane per hour and its malfunction may lead to major losses, therefore, an effective maintenance approach is of main interesting \cite{ELLANetwork}.

Reliability-centered maintenance consists of determining the most effective maintenance approach \cite{moubray1997reliability}. This process was firstly developed in the aviation industry for deciding what maintenance work is needed to keep aircraft airborne, driven by the need to improve reliability while containing the cost of maintenance \cite{sriram2003optimization}. Reliability analysis, can be used to estimate time-related parameters to the next machine stop \cite{lawless2011statistical}, providing information to manage and control of the preventive maintenance of harvesters which could create a production improvement and has potential for cost savings.

In reliability, common procedures are usually based on the assumption that the data follows a Weibull distribution. Introduced by Waloddi Weibull \cite{weibull1951wide} this distribution has convenient mathematical properties and its physiological failure process arise in many areas (see, Manton and Yashin \cite{manton2006inequalities}). However, this distribution cannot be used to describe data with non-monotone hazard function (bathtub, upside-down bathtub, to list a few). to overcome this problem many generalizations of the standard Weibull distribution have been proposed. For instance, Lai \cite{lai2014generalized} reviewed more than 20 generalizations of the Weibull distribution. 

In this paper, we considered five important generalized Weibull distributions with three parameters to describe the lifetime of two important components of the harvest sugarcane machines. The distributions considered are the Gamma-Weibull distribution \cite{stacy1962}, generalized Weibull (GW) distribution \cite{mudholkar1996generalization}, exponentiated Weibull (EW) distribution \cite{mudholkar1995exponentiated}, Marshall-Olkin Weibull (MOW) distribution \cite{marshall1997new} and the extended Poisson-Weibull (EPW) distribution \cite{ramos2018}. For each distribution, the mathematical background is presented and the parameters estimators are presented using the maximum likelihood estimators. Further, different discrimination procedures are used to obtain the best fit for each component. At the end, we propose a maintenance scheduling for the components of the harvesters using predictive analysis.

This work is divided as follows. Section 2 presents the literature review related the survival models adopted. Sections 3 exposes the data collection, and empirical analysis, as well as, carry out the predictive analysis based on the parametric models.  Finally, in Section 4, we present some final remarks related the contribution this study.

\section{Theoretical Background} 

In this section, we present the statistical background on the adopted distributions and its parameter estimation procedures. The following distributions are considered: Gamma-Weibull, generalized Weibull, exponentiated Weibull, Marshall-Olkin-Weibull and Marshall-Olkin-Weibull. Their choice is based on their flexibility to accommodate lifetime dataset with hazard functions with different shapes, for instance, constant, increasing, decreasing, bathtub and upside-down bathtub.

\subsection{The Gamma-Weibull distribution}

Introduced by Stacy \cite{stacy1962} the Gamma-Weibull distribution with three parameters is a flexible model for reliability data due to its ability to accommodates various forms of the hazard function. This distribution is also know as generalized gamma (GG) distribution as also generalize the two parameter gamma distribution, hereafter, we will refer this model as GG distribution to avoid confusion with the GW distribution. A random variable has GG distribution if its PDF is given by
\begin{equation}\label{denspgg}
f(t|\phi,\mu,\alpha)= \frac{\alpha}{\Gamma(\phi)}\mu^{\alpha\phi}t^{\alpha\phi-1}e^{-(\mu t)^{\alpha}} , \quad t>0 ,
\end{equation}
where $\alpha>0$, $\phi >0$ and $\mu >0$. The mean and variance of GG is given by
\begin{equation*}
E(X)=\frac{\Gamma\left(\phi + \frac{1}{\alpha}\right)}{\mu\Gamma(\phi)} \ \ \mbox{ and } \ \ V(X)=\frac{1}{\mu^2}\left\{\frac{\Gamma\left(\phi + \frac{2}{\alpha}\right)}{\Gamma(\phi)}-\left(\frac{\Gamma\left(\phi + \frac{1}{\alpha}\right)}{\Gamma(\phi)}\right)^2\right\},
\end{equation*}

Some relevant distributions are special cases such as the Weibull distribution (when $\phi=1$), the distribution Gamma ($\alpha=1$), Log-Normal (case limit when $\phi\rightarrow\infty$) and the Generalized Normal distribution ($\alpha=2$). As an example, the Generalized Normal distribution is also a distribution that includes several distributions known as, half-normal ($\phi=1/2, \mu=1/\sqrt{2}\sigma$), Rayleigh ($\phi=1,\mu=1/\sqrt{2}\sigma$), Maxwell-Boltzmann ($\phi=3/2$) e chi ($\phi=k/2 , k=1,2,\ldots$). The cumulative distribution function (CDF) is given by
\begin{equation*}
F(t|\phi,\mu,\alpha)= \int_{0}^{(\mu t)^{\alpha}}{\frac{1}{\Gamma(\phi)}w^{\phi -1}e^{-w}}dw =\frac{\gamma\left[\phi,(\mu t)^{\alpha}\right]}{\Gamma(\phi)}\ ,
\end{equation*}
where $\gamma[y,x]=\int_{0}^{x}{w^{y-1}e^{-w}}dw$ is the lower incomplete gamma function. The survival function is 
\begin{equation}\label{survivalgg}
S(t|\phi,\mu,\alpha)= 1-F(t|\phi,\mu,\alpha)=\frac{\Gamma\left[\phi,(\mu t)^{\alpha}\right]}{\Gamma(\phi)}\,,
\end{equation}
where $\Gamma[y,x]=\int_{x}^{\infty}{w^{y-1}e^{-w}}dw$ is the upper incomplete gamma function.

The hazard function of the GG distribution is
\begin{equation*}
h(t|\phi,\mu,\alpha))=\frac{f(t|\phi,\mu,\alpha)}{S(t|\phi,\mu,\alpha)}=\frac{\alpha\mu^{\alpha\phi}t^{\alpha\phi-1}\exp\left(-(\mu t)^{\alpha}\right)}{\Gamma\left[\phi,(\mu t)^{\alpha}\right]} .
\end{equation*}

This model is very flexible to describe lifetime data since it has the hazard function with constant, increasing, decreasing, bathtub and upside-down bathtub hazard rate.

For parameter estimation, let $T_1,\ldots,T_n$ be a random sample of size n, where $T\sim \f {GG}(\alpha,\mu,\phi)$. Then, the likelihood function related to the PDF (\ref{denspgg}) is given by 
\begin{equation}\label{verogg1} 
L(\phi,\mu,\alpha;\boldsymbol{t})=\frac{\alpha^n}{\f \Gamma(\phi)^n}\mu^{n\alpha\phi}\left\{\prod_{i=1}^n{t_i^{\alpha\phi-1}}\right\}\exp\left\{-\mu^{\alpha}\sum_{i=1}^n t_i^\alpha\right\}. \end{equation}

The log-likelihood is given by
\begin{equation*} 
l(\phi,\mu,\alpha;\boldsymbol{t})=n\log(\alpha)-n\log\Gamma(\phi)+n\alpha\phi\log(\mu)+(\alpha\phi-1)\sum_{i=1}^{n}\log(t_i)-\mu^{\alpha}\sum_{i=1}^n t_i^\alpha \,.\end{equation*}

Setting the partial derivatives $\frac{\partial}{\partial \alpha}\log(l(\phi,\mu,\alpha;\boldsymbol{t}))$, $\frac{\partial}{\partial \mu}\log(l(\phi,\mu,\alpha;\boldsymbol{t}))$ and $\frac{\partial}{\partial \phi}\log(l(\phi,\mu,\alpha;\boldsymbol{t}))$ equal to $0$, we obtain the following maximum likelihood estimators
\begin{equation*} 
\hat\mu=\left(\frac{1}{\hat\alpha}\frac{n}{\sum_{i=1}^n{t_i}^{\hat\alpha}\log(t_i)-\frac{\sum_{i=1}^n{t_i}^{\hat\alpha}}{n}\sum_{i=1}^n\log(t_i)}\right)^{\tfrac{1}{\alpha}},
\end{equation*}
\begin{equation*}
\hat\phi=\left(\frac{1}{\hat\alpha}\frac{\sum_{i=1}^n{t_i}^{\hat\alpha}}{\sum_{i=1}^n{t_i}^{\hat\alpha}\log(t_i)-\frac{\sum_{i=1}^n{t_i}^{\hat\alpha}}{n}\sum_{i=1}^n\log(t_i)}\right) \ \ \mbox{and}
\end{equation*}
\begin{equation*} n\hat\alpha\log(\hat\mu)+\hat{\alpha}\sum_{i=1}^n\log(t_i)-n \psi(\hat\phi)=0 ,\end{equation*}
the solution provides the maximum likelihood estimates (see, for instance, Ramos et. al \cite{ramos2014metodo,ramos2017bayesian}, Achcar et al. \cite{achcar2017some}). 

Under mild conditions that in some cases are not fulfill the estimators become unbiased for large samples and asymptotically efficient. Moreover, such estimators have asymptotically normal joint distribution given by
\begin{equation*} (\hat\phi,\hat\mu,\hat\alpha) \sim N_3[(\phi,\mu,\alpha),I^{-1}(\phi,\mu,\alpha)] \mbox{ for } n \to \infty , \end{equation*}
where $I(\boldsymbol{\theta})$ the Fisher information matrix is
\begin{equation*}
I(\alpha,\mu,\phi)=
\begin{bmatrix}
 \dfrac{1+2\psi(\phi)+\phi\psi ' (\phi)+\phi\psi(\phi)^2}{\alpha^2} & -\dfrac{1+\phi\psi(\phi)}{\mu} & -\dfrac{\psi(\phi)}{\alpha} \\
 -\dfrac{1+\phi\psi(\phi)}{\mu} & \dfrac{\phi\alpha^2}{\mu^2}  & \dfrac{\alpha}{\mu} \\
 -\dfrac{\psi(\phi)}{\alpha} & \dfrac{\alpha}{\mu} & \psi ' (\phi)
\end{bmatrix} .
\end{equation*}

\subsection{The generalized Weibull distribution}

Introduced by Mudholkar et al. \cite{mudholkar1996generalization} the generalized Weibull distribution has PDF given by 
\begin{equation}\label{densagnwei}
f(t|\lambda,\phi,\alpha)= {(\alpha\phi)}^{-1}{(t/\phi)}^{1/\alpha-1}{(1-\lambda{(t/\phi)}^{1/\alpha} )}^{1/\lambda-1},
\end{equation}
where $\lambda\in\mathbb{R}, \phi>0$ and $\alpha>0$. The CDF and the survival function are respectively given by
\begin{equation}\label{cdensagnwei}
F(t|\lambda,\phi,\alpha)= 1-{(1-\lambda{(t/\phi)}^{1/\alpha} )}^{1/\lambda} \ \ \mbox{ and } \ \ S(t|\lambda,\phi,\alpha)= {(1-\lambda{(t/\phi)}^{1/\alpha} )}^{1/\lambda}.
\end{equation}

The hazard function of the GW distribution is
\begin{equation*}
h(t|\lambda,\phi,\alpha)=\frac{(t/\phi)^{1/\alpha-1}}{\alpha\phi\left(1-\lambda(t/\phi)^{1/\alpha}\right)} \cdot
\end{equation*}

This model is very flexible to describe lifetime data since it has the hazard function with constant, increasing, decreasing, bathtub and upside-down bathtub hazard rate. The quantile function of the GW distribution has closed form and is given by 
\begin{equation*}
Q(u|\lambda,\phi,\alpha)=
\begin{cases}
\phi\left(-\log(1-u) \right)^\alpha & \text{ if } \ \lambda= 0,\\
\phi\left(\dfrac{1-(1-u)^\lambda}{\lambda} \right)^\alpha & \text{ if } \ \lambda\neq0 .
\end{cases}
\end{equation*}

For parameter estimation, let $T_1,\ldots,T_n$ be a random sample of size n, where $T\sim \f {GW}(\alpha,\mu,\phi)$. Then, the likelihood function related to the PDF (\ref{densagnwei}) is given by 
\begin{equation}\label{verogwa1} 
L(\lambda,\phi,\alpha;\boldsymbol{t})={(\alpha\phi)}^{-n}\prod_{i=1}^{n}{\left(\frac{t_i}{\phi
}\right)^{\frac{1}{\alpha}-1}{\left(1-\lambda{\left(\frac{t_i}{\phi
}\right)}^{1/\alpha} \right)}^{\frac{1}{\lambda}-1}}. \end{equation}

The log-likelihood is given by
\begin{equation}\label{logvergw}
\begin{aligned}
l(\lambda,\phi,\alpha;\boldsymbol{t})=&\left(\frac{1}{\lambda}-1\right)\sum_{i=1}^{n}\log\left(1-\lambda{\left(\frac{t_i}{\phi
}\right)}^{1/\alpha} \right) +\left(\frac{1}{\alpha}-1\right)\sum_{i=1}^{n}\log\left(\frac{t_i}{\phi}\right)-n\log(\alpha\phi)
\end{aligned}
\end{equation}

Setting the partial derivatives equal to $0$, we obtain the maximum likelihood estimators. Here, we following Mudholkar et al. \cite{mudholkar1996generalization} which consider the direct maximization of (\ref{logvergw}). Under mild conditions the obtained estimators are consistent and efficient with an asymptotically normal joint distribution given by
\begin{equation*} \boldsymbol{\hat{\Theta}} \sim N_3[\boldsymbol{\Theta},I^{-1}(\boldsymbol{\Theta})] \mbox{ as } n \to \infty , \end{equation*}
where $I(\boldsymbol{\Theta})$ is the $3\times 3$ Fisher information matrix associated to the vector of parameters $\boldsymbol{\Theta})$ and $I_{ij}(\boldsymbol{\Theta})$ is the Fisher information elements in $i$ and $j$ given by
\begin{equation*}\label{fisherinf}
I_{ij}(\boldsymbol{\Theta})=E\left[-\frac{\partial^2}{\partial \Theta_i \partial \Theta_j}l(\boldsymbol{\Theta};\mathcal{D})^2\right],\ i,j=1,2,3.
\end{equation*}

Since it is the the Fisher information matrix does not have closed-form expression for some terms, an alternative is to consider the observed information matrix, where the terms is given by
\begin{equation*}
H_{ij}(\boldsymbol{\Theta})=-\frac{\partial^2}{\partial \Theta_i \partial \Theta_j}l(\boldsymbol{\Theta};\boldsymbol{t})^2,\ i,j=1,2,3.
\end{equation*}

Hereafter, we considered the same approach to obtain the confidence intervals for the parameters from other distributions.

\subsection{The exponentiated Weibull distribution}

Introduced by Mudholkar et al. \cite{mudholkar1995exponentiated} the exponentiated Weibull distribution with PDF given by  
\begin{equation}\label{densexpw}
f(t|\sigma,\phi,\alpha)= \alpha\phi{\sigma}^{-1}{(t/\sigma)}^{\alpha-1}\exp\left(-{(t/\sigma)}^{\alpha} \right) \left(1-\exp\left(-{(t/\sigma)}^{\alpha} \right) \right)^{\phi-1},
\end{equation}
where $\sigma>0, \phi>0$ and $\alpha>0$.

The exponentiated Weibull distribution includes the Weibull distribution ($\phi=1$) and the exponentiated exponential distribution ($\alpha=1$). The survival function is given by
\begin{equation*}
S(t|\sigma,\phi,\alpha)=1-\left(1-\exp\left(-{(t/\sigma)}^{\alpha} \right) \right)^{\phi} .
\end{equation*}

The hazard function of the GG distribution is
\begin{equation}\label{fusobew}
h(t|\phi,\mu,\alpha)=\frac{\alpha\phi{(t/\sigma)}^{\alpha-1}\exp\left(-{(t/\sigma)}^{\alpha} \right) \left(1-\exp\left(-{(t/\sigma)}^{\alpha} \right) \right)^{\phi-1}}{\sigma\left(1-\left(1-\exp\left(-{(t/\sigma)}^{\alpha} \right) \right)^{\phi}\right)} \cdot
\end{equation}

This model is very flexible to describe lifetime data since it has the hazard function with constant, increasing, decreasing, bathtub and upside-down bathtub hazard rate. Additionally, the quantile function of the EW distribution has closed form and is given by 
\begin{equation*}
Q(u|\sigma,\phi,\alpha)=\sigma\left(-\log\left(1-u^{1/\phi}\right)\right)^{1/\alpha}.
\end{equation*}

The $k$-th moment of the EW distribution is given by
\begin{equation*}
\begin{aligned}
\mu_k&=\int_{0}^{1} Q(u|\sigma,\phi,\alpha)^kdu=\theta\sigma^k\Gamma\left(\frac{k}{\alpha}+1\right) \left(1+\sum_{i=1}^{\infty}a_i\left[(i+1)^{k/\alpha+1}\right]\right), \ k\in\mathbb{N},
\end{aligned}
\end{equation*}
where $a_i=(-1)^{i}(\theta-1)(\theta-2)\ldots\left(\theta-1-\overline{i-1}\right)(i\,!)^{-1}$. The proof of this equality is presented by Choudhury \cite{choudhury2005simple}.

For parameter estimation, let $T_1,\ldots,T_n$ be a random sample of size n, where $T\sim \f{EW}(\sigma,\phi,\alpha)$. Then, the likelihood function related to the PDF (\ref{densexpw}) is given by 
\begin{equation}\label{veroexw1} 
L(\sigma,\phi,\alpha;\boldsymbol{t})=\frac{\alpha^n\phi^n}{{\sigma}^{n}}\prod_{i=1}^n{\left(\frac{t_i}{\sigma}\right)}^{\alpha-1}\left(1-\exp\left(-{\left(\frac{t_i}{\sigma}\right)}^{\alpha} \right) \right)^{\phi-1}\exp\left(-\sum_{i=1}^n{\left(\frac{t_i}{\sigma}\right)}^{\alpha} \right) . \end{equation}

The log-likelihood is given by
\begin{equation*}
\begin{aligned}
l(\sigma,\phi,\alpha;\boldsymbol{t})=& \, n\log(\alpha\phi)-n\alpha\log(\sigma)+(\alpha-1)\sum_{i=1}^n\log(t_i)-\sum_{i=1}^n{\left(\frac{t_i}{\sigma}\right)}^{\alpha} \\ & + (\phi-1)\sum_{i=1}^{n}\log\left(1-\exp\left(-{\left(\frac{t_i}{\sigma}\right)}^{\alpha} \right) \right).
\end{aligned}
\end{equation*}

Setting the partial derivatives $\frac{\partial}{\partial \sigma}l(\sigma,\phi,\alpha;\boldsymbol{t})$, $\frac{\partial}{\partial \phi}l(\sigma,\phi,\alpha;\boldsymbol{t})$ and $\frac{\partial}{\partial \alpha}l(\sigma,\phi,\alpha;\boldsymbol{t})$ equal to $0$, we obtain the following maximum likelihood estimators
{\small
\begin{equation*} 
\frac{n}{\alpha}-n\log(\sigma)+\sum_{i=1}^n\log(t_i)+\frac{(\phi-1)}{\sigma^\alpha}\sum_{i=1}^{n}\frac{t_i^{\alpha}\log\left(t_i/\sigma\right)}{\exp\left({\left(\frac{t_i}{\sigma}\right)}^{\alpha} \right)-1}-\sum_{i=1}^{n}{\left(\frac{t_i}{\sigma}\right)}^{\alpha}\log\left(\frac{t_i}{\sigma}\right)\exp\left(-{\left(\frac{t_i}{\sigma}\right)}^{\alpha} \right)=0,
\end{equation*} }
\begin{equation*}  
-\frac{n\alpha}{\sigma}-\frac{\alpha}{\sigma^{\alpha+1}}\sum_{i=1}^n{t_i^{\alpha}}+ \frac{\alpha}{\sigma^\alpha}\sum_{i=1}^{n}\frac{(\phi-1)t_i^{\alpha}}{\sigma-\sigma\exp\left({\left(\frac{t_i}{\sigma}\right)}^{\alpha} \right) }=0 , \ \ \mbox{ where
 }
\end{equation*}
\begin{equation*} 
\phi=-\frac{n}{\sum_{i=1}^{n}\log\left(1-\exp\left(-{\left(\frac{t_i}{\sigma}\right)}^{\alpha} \right) \right)} \cdot
\end{equation*}

\subsection{The Marshall-Olkin-Weibull distribution}

Marshall and Olkin \cite{marshall1997new} introduced a new procedure for adding a new parameter into a family of distribution. In this case, the authors applied such procedure in the Weibull distribution. The obtained PDF of the MOW distribution is given by
\begin{equation}\label{fdpmow} 
f(t;\lambda,\alpha,\gamma)= \frac{\alpha\gamma\lambda t^{\gamma-1}e^{-\lambda t^\gamma}}{\left(1-(1-\alpha)e^{-\lambda t^\gamma}\right)^2} ,
\end{equation}
where $\lambda>0$, $\alpha>0$ and $\gamma>0$. Cordeiro and Lemonte \cite{cordeiro2013marshall} derived many properties and the parameter estimators for the MOW distribution, the following results were obtained from the cited work. The survival function is given by
\begin{equation*}
S(t|\lambda,\alpha,\gamma)=1-\frac{1-e^{-\lambda t^\gamma}}{1-(1-\alpha)e^{-\lambda t^\gamma}} \cdot
\end{equation*}

The hazard function of the MOW distribution is
\begin{equation}
h(t;\lambda,\alpha,\gamma)=\frac{\gamma\lambda t^{\gamma-1}}{1-(1-\alpha)e^{-\lambda t^\gamma}} \cdot
\end{equation}

This model is very flexible to describe lifetime data since it has the hazard function with constant, increasing, decreasing, bathtub and upside-down bathtub hazard rate. Additionally, the quantile function of the MOW distribution has closed form and is given by 
\begin{equation*}
Q(u|\lambda,\alpha,\gamma)=\lambda^{-1/\gamma}\left(\log\left(\frac{1-(1-\alpha)u}{1-u}\right)\right)^{1/\gamma}.
\end{equation*}

For parameter estimation, let $T_1,\ldots,T_n$ be a random sample of size n, where $T\sim \f{MOW}(\lambda,\alpha,\gamma)$. Then, the likelihood function related to the PDF (\ref{fdpmow}) is given by 
\begin{equation}
L(\lambda,\alpha,\gamma;\boldsymbol{t})=\alpha^n\gamma^n\lambda^n\prod_{i=1}^{n}\frac{ t_i^{\gamma-1}}{\left(1-(1-\alpha)e^{-\lambda t_i^\gamma}\right)^2}\exp\left(-\lambda\sum_{i=1}^{n}t_i^\gamma\right) . \end{equation}

The log-likelihood is given by
\begin{equation*} 
\begin{aligned}
l(\lambda,\alpha,\gamma;\boldsymbol{t})=& \, n\log(\alpha)+ n\log(\gamma)+ n\log(\lambda)+(\gamma-1)\sum_{i=1}^{n}\log(t_i)-\lambda\sum_{i=1}^{n}t_i^\gamma \\ & -2\sum_{i=1}^{n}\log\left(1-(1-\alpha)e^{-\lambda t_i^\gamma}\right).
\end{aligned}
\end{equation*}

Setting the partial derivatives $\frac{\partial}{\partial \lambda}l(\lambda,\alpha,\gamma;\boldsymbol{t})$, $\frac{\partial}{\partial \alpha}l(\lambda,\alpha,\gamma;\boldsymbol{t})$ and $\frac{\partial}{\partial \gamma}l(\lambda,\alpha,\gamma;\boldsymbol{t})$ equal to $0$, we obtain the following maximum likelihood estimators
\begin{equation*}  
\frac{n}{\alpha}-2\sum_{i=1}^n\frac{e^{-\lambda t_i^\gamma}}{1-(1-\alpha)e^{-\lambda t_i^\gamma}}=0,
\end{equation*}
\begin{equation*}
\frac{n}{\lambda}-\sum_{i=1}^n t_i^\gamma-2(1-\alpha)\sum_{i=1}^n\frac{t_i^\gamma e^{-\lambda t_i^\gamma}}{1-(1-\alpha)e^{-\lambda t_i^\gamma}}=0 \quad \mbox{and}
\end{equation*} 
\begin{equation*} 
\frac{n}{\gamma}+\sum_{i=1}^n\log(t_i) -\lambda\sum_{i=1}^nt_i^\gamma\log(t_i)  -2(1-\alpha)\lambda\sum_{i=1}^n\frac{t_i^\gamma\log(t_i) e^{-\lambda t_i^\gamma}}{1-(1-\alpha)e^{-\lambda t_i^\gamma}}=0,
\end{equation*}
for more details, see Cordeiro and Lemonte \cite{cordeiro2013marshall}.

\subsection{The extended Poisson-Weibull distribution}

Ramos et al. \cite{ramos2018} introduced the extended Poisson-Weibull (EPW) distribution as a generalization of Weibull-Poisson distribution (see Hemmati et al. \cite{hemmati2011new}) where its PDF is given by
\begin{equation}\label{fdpweib} 
f(t;\lambda,\alpha,\beta)= \frac{\alpha\lambda\beta  t^{\alpha-1} e^{-\beta t^\alpha-\lambda e^{-\beta t^\alpha}}}{1 - e^{-\lambda}},
\end{equation}
where $\lambda\in\mathbb R^*$, $\beta>0$ and $\alpha>0$. The survival function is given by
\begin{equation*}
S(t|\lambda,\alpha,\beta)=\frac{1-\exp\left({-\lambda e^{-\beta t^\alpha}}\right)}{1-e^{-\lambda}}.
\end{equation*}

The hazard function of the GG distribution is
\begin{equation*}
h(t;\lambda,\alpha,\beta)=\lambda \beta t^{\alpha-1}e^{-\beta t^\alpha -\lambda e^{-\beta t^\alpha}}\left(1-e^{-\lambda e^{-\beta t^\alpha}}\right)^{-1}.
\end{equation*}

This model is very flexible to describe lifetime data since it has the hazard function with constant, increasing, decreasing, bathtub and upside-down bathtub hazard rate. Additionally, the quantile function of the EPW distribution has closed form and is given by 
\begin{equation*}
Q(u|\lambda,\alpha,\beta)=\left(-\frac{1}{\beta}\log\left(1-\frac{\log\left((e^{\lambda} - 1)p + 1\right)}{\lambda}\right)\right)^{1/\alpha}.
\end{equation*}

For parameter estimation, let $T_1,\ldots,T_n$ be a random sample of size n, where $T\sim \f{EPW}(\lambda,\alpha,\beta)$. Then, the likelihood function related to the PDF (\ref{densexpw}) is given by 
\begin{equation*}
L(\lambda,\alpha,\beta;\boldsymbol{t})=\frac{\alpha^n\lambda^n\beta^n}{\left(1 - e^{-\lambda}\right)^n} \prod_{i=1}^{n} t_i^{\alpha-1} \exp\left(-\beta \sum_{i=1}^{n} t_i^\alpha-\lambda\sum_{i=1}^{n}e^{-\beta t_i^\alpha}\right). \end{equation*}

The log-likelihood is given by
\begin{equation*}
\begin{aligned}
l(\lambda,\alpha,\beta;\boldsymbol{t})=& \, n\log(\alpha\lambda\beta)-n\log\left(1 - e^{-\lambda}\right)+(\alpha-1)\sum_{i=1}^{n}\log(t_i)-\beta \sum_{i=1}^{n} t_i^\alpha-\lambda\sum_{i=1}^{n}e^{-\beta t_i^\alpha}.
\end{aligned}
\end{equation*}

Setting the partial derivatives $\frac{\partial}{\partial \lambda}l(\lambda,\alpha,\beta;\boldsymbol{t})$, $\frac{\partial}{\partial \alpha}l(\lambda,\alpha,\beta;\boldsymbol{t})$ and $\frac{\partial}{\partial \beta}l(\lambda,\alpha,\beta;\boldsymbol{t})$ equal to $0$, we obtain the following maximum likelihood estimators
\begin{equation*}  
\frac{n}{\lambda}+\frac{n}{1-e^\lambda}-\sum_{i=1}^{n}e^{-\beta t_i^\alpha}=0,
\end{equation*}
{\small
\begin{equation*} 
\frac{n}{\alpha}+\sum_{i=1}^{n}\log(t_i)-\beta \sum_{i=1}^{n} t_i^\alpha\log(t_i)+\beta\lambda\sum_{i=1}^{n}t_i^\alpha\log(t_i)e^{-\beta t_i^\alpha}=0
\end{equation*} }
and
\begin{equation*} 
\frac{n}{\beta}-\sum_{i=1}^{n} t_i^\alpha+\lambda\sum_{i=1}^{n}t_i^\alpha e^{-\beta t_i^\alpha}=0 .
\end{equation*}

\subsection{Goodness of fit}

Firstly, in order to verify the behavior of the empirical data the TTT-plot (total time on test) was considered (Barlow and Campo \cite{barlow1975total}). The TTT-plot is obtained through the plot of $[r/n,G(r/n)]$ where 
\begin{equation*}
G(r/n)= \left(\sum_{i=1}^{r}t_i +(n-r)t_{(r)}\right)/{\sum_{i=1}^{n}t_i}, \quad r=1,\ldots,n, \ \ i=1,\ldots,n  ,
\end{equation*}
and $t_{(i)}, i=1,\cdots,n$ is the ordered data. For data with concave (convex) curve , the hazard function has increasing (decreasing) shape. In the case of the behavior starts convex and then becomes concave (concave and then convex) the hazard function has bathtub (inverse bathtub) shape. 

The goodness of fit is checked considering the Kolmogorov-Smirnov (KS) test. This procedure is based on the KS statistic $D_n=\sup\left\vert F_n(t)-F(t;\phi ,\lambda ,\alpha)\right\vert$, where $\sup t$ is the supremum of the set of distances, $F_n(t)$ is the empirical distribution function and $F(t;\alpha ,\beta ,\lambda)$ is c.d.f. A hypothesis test is conducted at the $5\%$ level of significance to test whether or not the data comes from $F(t;\alpha ,\beta ,\lambda)$. In this case, the null hypothesis is rejected if the returned p-value is smaller than $0.05$.

The following discrimination criterion methods were adopted: Akaike information criteria (AIC) and the corrected AIC (AICc) computed respectively by $AIC=-2l(\hat{\boldsymbol{\theta}};\boldsymbol{t})+2k$ and $AICc=AIC+2\,k\,(k+1){(n-k-1)}^{-1}$, where $k$ is the number of parameters to be fitted and $\hat{\boldsymbol{\theta}}$ is MLEs of $\boldsymbol{\theta}$. For a set of candidate models for $\boldsymbol{t}$, the best one provides the minimum values. 

\section{Data collection and Empirical Analysis}

The dataset came from two sources: a manual stop system, which brings the history of revisions and corrective stops of two sugarcane harvesters; and data from the onboard computers of the harvesters, which provide information on the operation of the machine. The data were collected from January 2015 to August 2017, a period corresponding to 2.5 harvests (crops), i.e., a period of thirty months of activity.

\subsection{Empirical Analysis}

Firstly, considering all the stops and their reasons, records of the performance of the predictive maintenance is needed to be observed. In total, 1347 stops were observed, which 186 were preventative and 1161 corrective stops. Thus, it is possible to observe the superior amount of unplanned stops, thus questioning the effectiveness of preventive maintenance. Table \ref{tab:records} shows the failure among the harvests, considering both machines analysis.

\begin{table}[htbp]
  \centering
  \caption{Maintenance Distribution preventive and corrective stops per crop.}
    \begin{tabular}{ccccccc}
    \toprule
    \multicolumn{1}{c}{\multirow{2}[4]{*}{}} & \multicolumn{2}{c}{Crop 1} & \multicolumn{2}{c}{Crop 2} & \multicolumn{2}{c}{Crop 3} \\
\cmidrule{2-7}    \multicolumn{1}{c}{} & \multicolumn{1}{p{4.215em}}{Preventive} & \multicolumn{1}{p{4.215em}}{Corrective} & \multicolumn{1}{p{4.215em}}{Preventive} & \multicolumn{1}{p{4.215em}}{Corrective} & \multicolumn{1}{p{4.215em}}{Preventive} & \multicolumn{1}{p{4.215em}}{Corrective} \\
    \midrule
    Machine A & 39 & 232 & 32 & 255 & 23 & 127 \\
    Machine B & 37 & 199 & 32 & 182 & 23 & 166 \\
    \bottomrule
    \end{tabular}%
  \label{tab:records}%
\end{table}%

\begin{figure}[!h]
\centering
\includegraphics[scale=0.7]{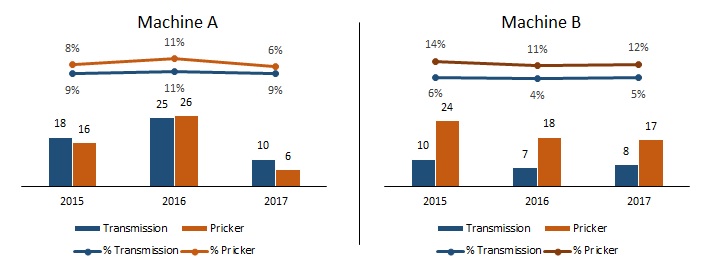}
\caption{Maintenance distribution in each harvester.}\label{graf2}
\end{figure}

The Pricker and Transmission from each machine were selected given their complexity in the maintenance. Figure \ref{graf2} describes the number of failures per year divided by harvest, considering their temporal sparsity, by which items analyzed in this report, correspond to 18\% of the stops.

It is possible to notice a  difference in the machines' behaviour, both machines appear to be equally affected by the problems of Transmission and Pricker, but the machine B is more affected by problems with the Pricker. Further, reliability models were individually adjusted, and thereby compared, as described in the next section.


\subsection{Preventive maintenance}

In this section, we discuss a parametric approach in order to perform a predictive analysis for the lifetime of the components.

\subsubsection{Pricker from machine A}

Table \ref{tablepicadora} presents a high defect rate after a short repair time as well, compromising the cost of the production. The experiment considered a total period of 30 months, as said before. Then operating equipment had three off-seasons, these periods were not included in the dataset. The equipment was only observed during the time of its active operation.

\begin{table}[!h]
	\caption{Dataset related to the sugarcane harvester's pricker.}
\centering 
	{\begin{tabular}{c c c c c c c c c c c c c c} 
		\hline 
1 & 1 & 1 & 1 & 1 & 1 & 1 & 1 & 2 & 2 & 2 & 2 \\
2 & 3 & 3 & 3 & 3 & 3 & 4 & 4 & 4 & 5 & 5 & 5 \\
6 & 6 & 7 & 8 & 9 & 11 & 11 & 12 & 14 & 16 & 18 & 18 \\
18 & 22 & 22 & 23 & 29 & 32 & 34 & 38 & 41 & 46 & 53 & 53 \\
 [0ex] 
		\hline 
	\end{tabular}}\label{tablepicadora}
\end{table}

\begin{figure}[!htb]
\centering
\includegraphics[scale=0.6]{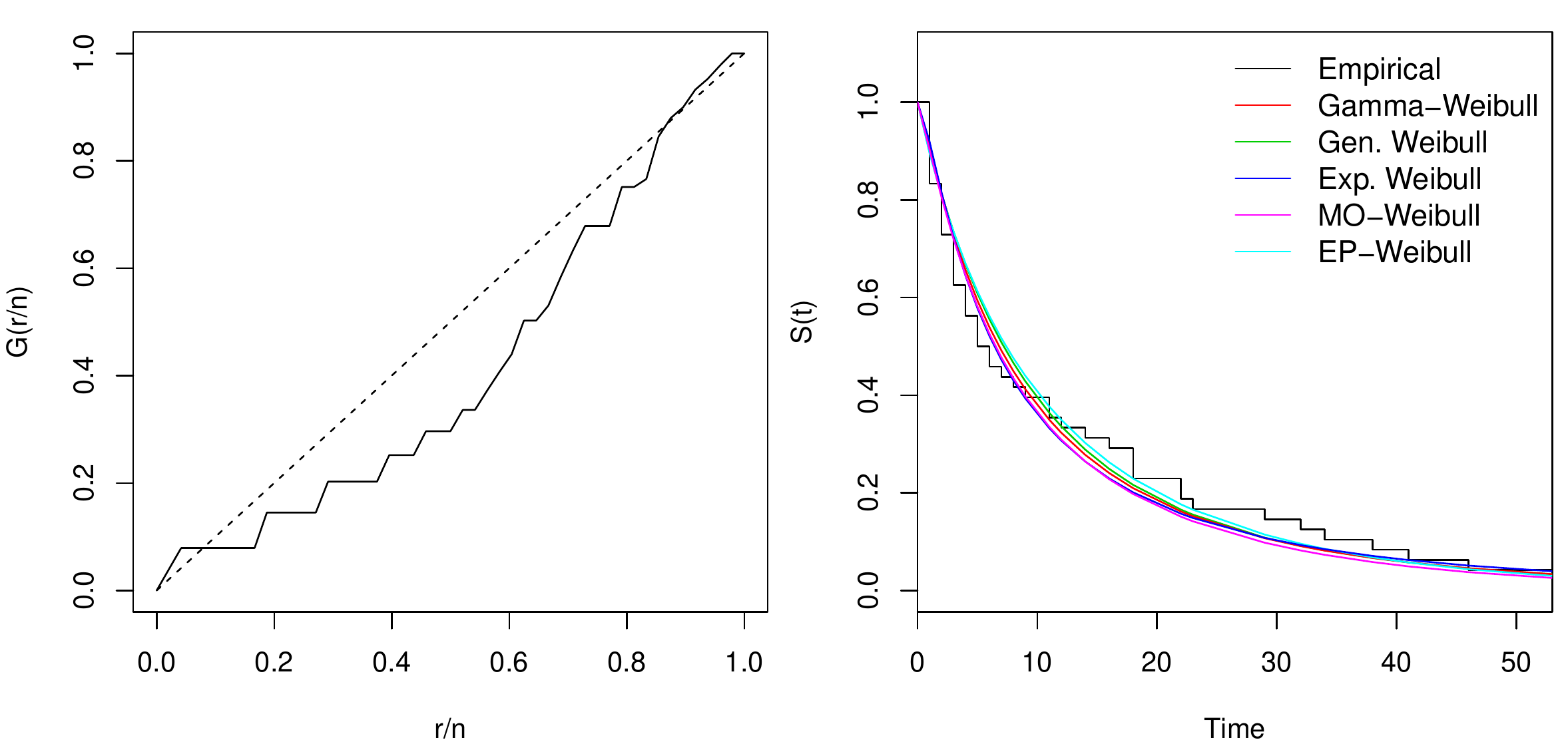}
\caption{Pricker A empirical (left-panel) TTT-plot, and (right-panel) contains the fitted survival superimposed to the empirical survival function and the hazard function adjusted by distribution}\label{graf-pic-a}
\end{figure}

Figure \ref{graf-pic-a} presents the TTT-plot and the survival function fitted by different generalizations of the Weibull distribution.

From the TTT-plot we observed that the proposed data has unimodal hazard rate, which implies that all the proposed models may be used to describe the proposed dataset. Table \ref{tableaic1} presents the results of AIC, and AICc in order to discriminate the best fit. 

\begin{table}[ht]
\caption{Results of AIC and AICc criteria and the p-value from the KS test for all fitted distributions considering the pricker A.}
\centering 
\begin{center}
  \begin{tabular}{ c | c | c | c | c | c }
    \hline
		Criteria & \ Gen. Gamma  \ & \ Gen. Weibull &  \ Exp. Weibull &  \ MO Weibull &  \ EPW\\ \hline
      AIC & 341.013 & 343.696 & \textbf{340.435} & 341.317 & 342.750 \\ \hline
			AICc & 335.559 & 338.241 & \textbf{334.981} & 335.862 & 343.296 \\ \hline
			KS   & 0.6735 & 0.6046 & 0.7447 & 0.7457 & 0.5751 \\ \hline
  \end{tabular}\label{tableaic1}
\end{center}
\end{table}

Among the proposed models the Exponentiated Weibull distribution has superior goodness of fit since the AIC and AICc returned smaller values. Therefore, using the Exponentiated Weibull distribution we computed the maximum likelihood estimates and the predictive value for $25\%$  (see Table \ref{reprepicadora}). Hereafter, as we considered the quantile function to obtain the predictive value, the confidence intervals related to this estimate were obtained from bootstrap technique (see Efron and Tibshirani \cite{efron1994introduction}).

\begin{table}[!h]
	\caption{MLEs, Standard deviations and  $95\%$ credibility intervals for $\alpha$, $\theta$, $\sigma$  and $y*$ related to the EW distribution}
	\centering 
{\small
	{\begin{tabular}{ c | c |  c| c }
			\hline
			$\boldsymbol{\theta}$ & MLE & SD & CI$_{95\%}(\boldsymbol{\theta})$ \\ \hline
			\ \ $\alpha$ \ \    & 0.379 &  0.044  &  (0.3181; 0.4969)  \\ \hline
			\ \ $\theta$  \ \   & 6.446 &  0.703  &  (4.7070; 7.6915)  \\ \hline
			\ \ $\sigma$  \ \   & 0.727 &  0.290  &  (0.3897; 1.4691)  \\ \hline
			\ \ $y*$   \ \        & 3.093 &  0.582  &  (1.7374; 4.0236)  \\ \hline
		\end{tabular}}\label{reprepicadora} }
\end{table}

From Table \ref{reprepicadora} we observe that the predictive maintenance should be done in approximately 3 days  after the last failure with confidence interval between 2 and 4 days.

\subsubsection{Pricker from machine B}

A similar behavior is observed for the Pricker in the machine B, shown in Table \ref{tablepicadorb} presenting a high defect rate as well. The approach was maintained considering only the time during its active operation.
\begin{table}[!h]
	\caption{Dataset related to the sugarcane harvester's pricker B.}
\centering 
	{\begin{tabular}{c c c c c c c c c c c c} 
		\hline 
1 & 1 & 1 & 1 & 1 & 1 & 1 & 1 & 1 & 1 & 1 & 1 \\
2 & 2 & 2 & 3 & 3 & 3 & 3 & 3 & 3 & 4 & 4 & 5 \\
5 & 5 & 5 & 5 & 5 & 5 & 6 & 7 & 7 & 8 & 8 & 8 \\
8 & 8 & 9 & 9 & 11 & 11 & 11 & 11 & 11 & 11 & 12 & 13 \\
14 & 16 & 16 & 21 & 23 & 24 & 27 & 28 & 38 & 43 & 44 \\
 [0ex] 
		\hline 
	\end{tabular}}\label{tablepicadorb}
\end{table}

\begin{figure}[!htb]
\centering
\includegraphics[scale=0.6]{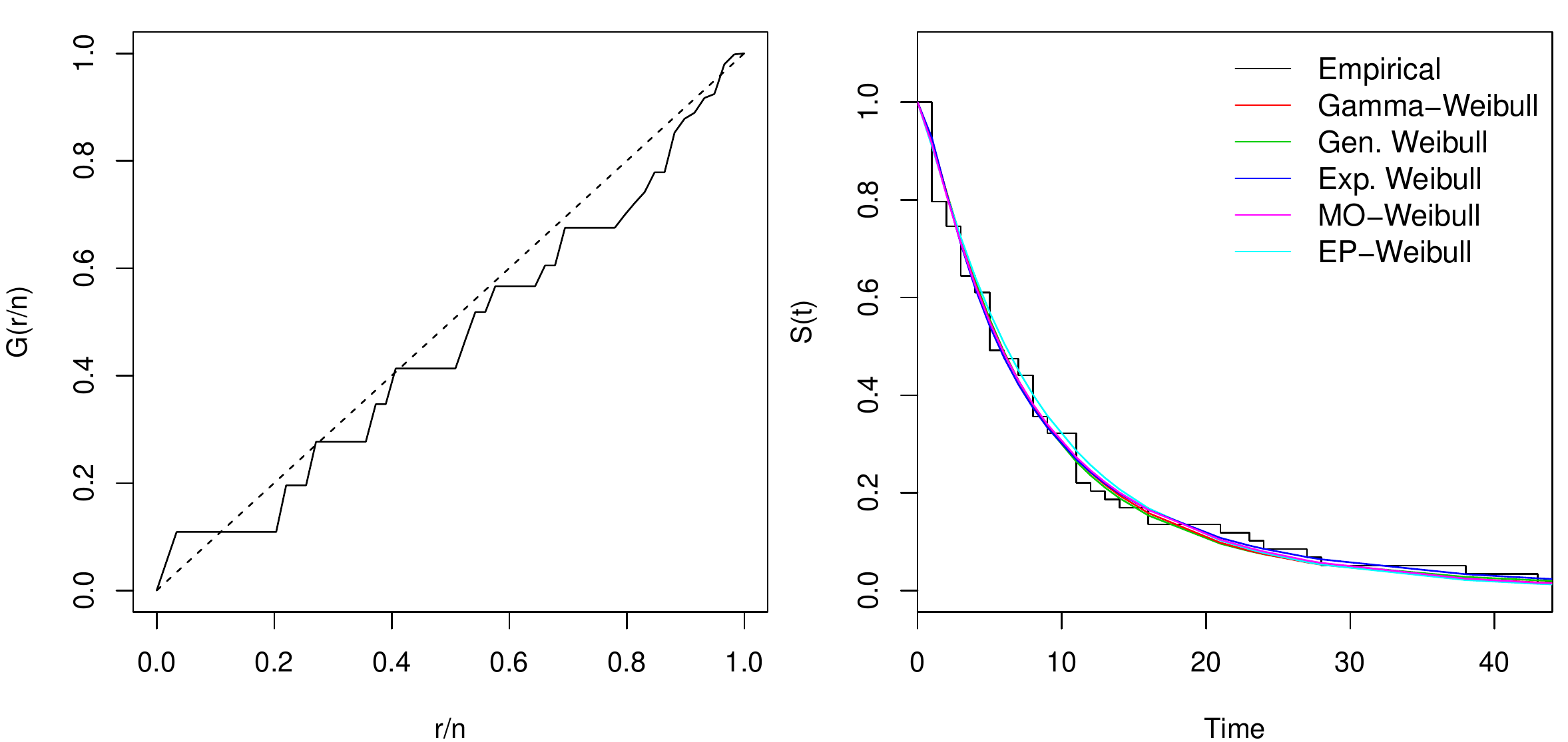}
\caption{Pricker B empirical (left-panel) TTT-plot, and (right-panel) contains the fitted survival superimposed to the empirical survival function and the hazard function adjusted by distribution}\label{graf-pic-b}
\end{figure}

Figure \ref{graf-pic-b} presents the TTT-plot and the survival function fitted by different generalizations of the Weibull distribution, similar to the previous machine.

From the TTT-plot we observed that the proposed data has unimodal hazard rate, which implies that all the proposed models may be used to describe the proposed dataset. Table \ref{tableaic2} presents the results of AIC, and AICc in order to discriminate the best fit with the EW distribution. In this case, we observe the similarity between both machines for this component. 
\begin{table}[ht]
\caption{Results of AIC and AICc criteria and the p-value from the KS test for all fitted distributions considering the pricker B.}
\centering 
\begin{center}
  \begin{tabular}{ c | c | c | c | c | c }
    \hline
		Criteria & \ Gen. Gamma  \ & \ Gen. Weibull &  \ Exp. Weibull &  \ MO Weibull &  \ EPW\\ \hline
      AIC & 382.063 & 384.102 & \textbf{381.790} & 382.641 & 383.772 \\ \hline
			AICc & 376.500 & 378.538 & \textbf{376.226} & 377.077 & 384.209 \\ \hline
			KS   & 0.3055 &  0.3628 & 0.2737  & 0.3900 & 0.4443 \\ \hline
  \end{tabular}\label{tableaic2}
\end{center}
\end{table}

Thus, the maximum likelihood estimates for the EW distribuiton were computed as well as the predictive value for $25\%$. Table \ref{reprepicadoraB} presents the MLEs, Standard deviations and $95\%$ credibility intervals for $\alpha$, $\theta$, $\sigma$  and $y*$ related to the EW distribution.

\begin{table}[!h]
	\caption{MLEs, Standard deviations and  $95\%$ credibility intervals for $\alpha$, $\theta$, $\sigma$  and $y*$ related to the EW distribution}
	\centering 
{\small
	{\begin{tabular}{ c | c |  c| c }
			\hline
			$\boldsymbol{\theta}$ & MLE & SD & CI$_{95\%}(\boldsymbol{\theta})$ \\ \hline
			\ \ $\alpha$ \ \    & 0.457 &  0.050  &  (0.3955; 0.5835)  \\ \hline
			\ \ $\theta$  \ \   & 5.434 &  0.763  &  (3.5493; 6.8379)  \\ \hline
			\ \ $\sigma$  \ \   & 1.083 &  0.327  &  (0.6879; 1.9727)  \\ \hline
			\ \ $y*$   \ \      & 2.497 &  0.459  &  (1.8212; 3.5760)  \\ \hline
		\end{tabular}}\label{reprepicadoraB} }
\end{table}

Table \ref{reprepicadoraB} results suggest that predictive maintenance should be done in approximately 3 days, considering a point estimation, or given a 95\% credibility interval would be between 2 to 4 days approximated. Thereby, Pricker among machines showed no difference performance so ever.

\subsubsection{Transmission from machine A}

Table \ref{tabletramissiona} shows that more than 50\% of the defect rate appears until 8 days right after its repair for the Transmission for the machine A. 

\begin{table}[!h]
	\caption{Dataset related to the sugarcane harvester's transmission A.}
\centering 
	{\begin{tabular}{c c c c c c c c c c c c} 
		\hline 
1 & 1 & 1 & 1 & 1 & 1 & 1 & 1 & 1 & 1 & 1 & 1 \\
2 & 2 & 2 & 3 & 3 & 3 & 3 & 4 & 5 & 6 & 6 & 6 \\
6 & 7 & 7 & 8 & 8 & 8 & 11 & 11 & 12 & 13 & 13 & 13 \\
14 & 15 & 16 & 17 & 18 & 18 & 19 & 19 & 21 & 22 & 23 & 29 \\
31 & 32 & 34 & 44 & 52 \\
 [0ex] 
		\hline 
	\end{tabular}}\label{tabletramissiona}
\end{table}

\begin{figure}[!htb]
\centering
\includegraphics[scale=0.6]{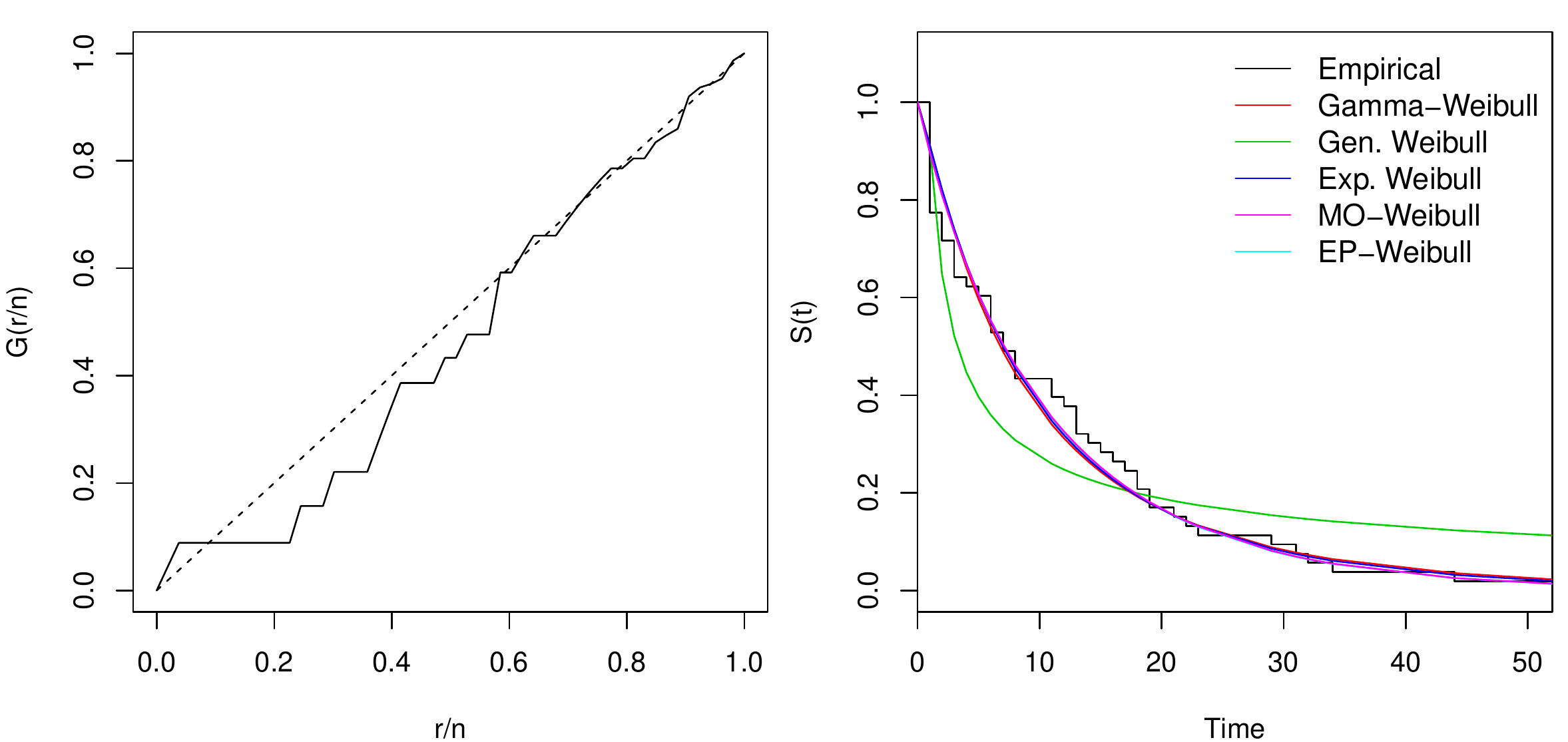}
\caption{Transmission A empirical (left-panel) TTT-plot, and (right-panel) contains the fitted survival superimposed to the empirical survival function and the hazard function adjusted by distribution}\label{graf-transmissor-a}
\end{figure}

Figure \ref{graf-transmissor-a} presents the TTT-plot and the survival function fitted by different generalizations of the Weibull distribution.

From the TTT-plot we observed that the proposed data has also fulfill the hazard rate shape preposterous to different generalizations of the Weibull distribution used. Table \ref{tableaic2ta} presents the results of AIC, and AICc in order to discriminate the best fit. 

\begin{table}[ht]
\caption{Results of AIC and AICc criteria and the p-value from the KS test for all fitted distributions considering the pricker B.}
\centering 
\begin{center}
  \begin{tabular}{ c | c | c | c | c | c }
    \hline
		Criteria & \ Gen. Gamma  \ & \ Gen. Weibull &  \ Exp. Weibull &  \ MO Weibull &  \ EWP\\ \hline
      AIC  & 368.074 & 381.036 & 368.271 & 368.375 & 368.385 \\ \hline
			AICc & 362.563 & 375.526 & 362.761 & 362.864 & 368.875 \\ \hline
			KS   & 0.2532 &  0.0035 & 0.2672  & 0.3945 & 0.3738 \\ \hline
  \end{tabular}\label{tableaic2ta}
\end{center}
\end{table}

As can be seen from the  Table \ref{tableaic2ta} the GW distribution has the p-value of the KS test smaller than 0.05, therefore, is not a possible candidate to fit the data. Overall, the GG distribution has a better fit since has the smaller AIC and AICc. Therefore, we computed the maximum likelihood estimates and the predictive value for $25\%$ using the GG distribution. Table \ref{repretransA} presents the MLEs, Standard deviations and $95\%$ credibility intervals for $\phi$, $\mu$, $\alpha$  and $y*$ related to the GG distribution.

\begin{table}[!h]
	\caption{MLE, Standard deviation and  $95\%$ credibility intervals for $\phi$, $\mu$, $\alpha$  and $y*$ related to the GG distribution}
	\centering 
{\small
	{\begin{tabular}{ c | c |  c| c }
			\hline
			$\boldsymbol{\theta}$ & MLE & SD & CI$_{95\%}(\boldsymbol{\theta})$ \\ \hline
			\ \ $\phi$ \ \    & 3.011 &  0.543  &  (1.7396; 3.9936)  \\ \hline
			\ \ $\mu$  \ \    & 1.086 &  0.525  &  (0.2389; 2.2682)  \\ \hline
			\ \ $\alpha$  \ \ & 0.495 &  0.075  &  (0.4214; 0.7124)  \\ \hline
			\ \ $y*$   \ \    & 2.807 &  0.635  &  (1.8487; 4.3526)  \\ \hline
		\end{tabular}}\label{repretransA} }
\end{table}

Table \ref{repretransA} results suggest that predictive maintenance should be done in approximately 3 days, considering a point estimation, or given a 95\% credibility interval would be between 2 to 4 days approximated.

\subsubsection{Transmission from machine B}

Comparing to the others equipments, the transmission from the machine B presented smaller number of occurrence. Table \ref{tabletramissionaB} shows the sparsity of the dataset related to the sugarcane harvester's transmission B.

\begin{table}[!h]
	\caption{Dataset related to the sugarcane harvester's transmission B.}
\centering 
	{\begin{tabular}{c c c c c c c c c c c c} 
		\hline 
1 & 2 & 3 & 3 & 4 & 5 & 6 & 6 & 7 & 9 & 11 & 12 \\
12 & 18 & 19 & 21 & 23 & 28 & 31 & 31 & 35 & 37 & 39 & 46 \\
61 \\
 [0ex] 
		\hline 
	\end{tabular}}\label{tabletramissionaB}
\end{table}

\begin{figure}[!htb]
\centering
\includegraphics[scale=0.6]{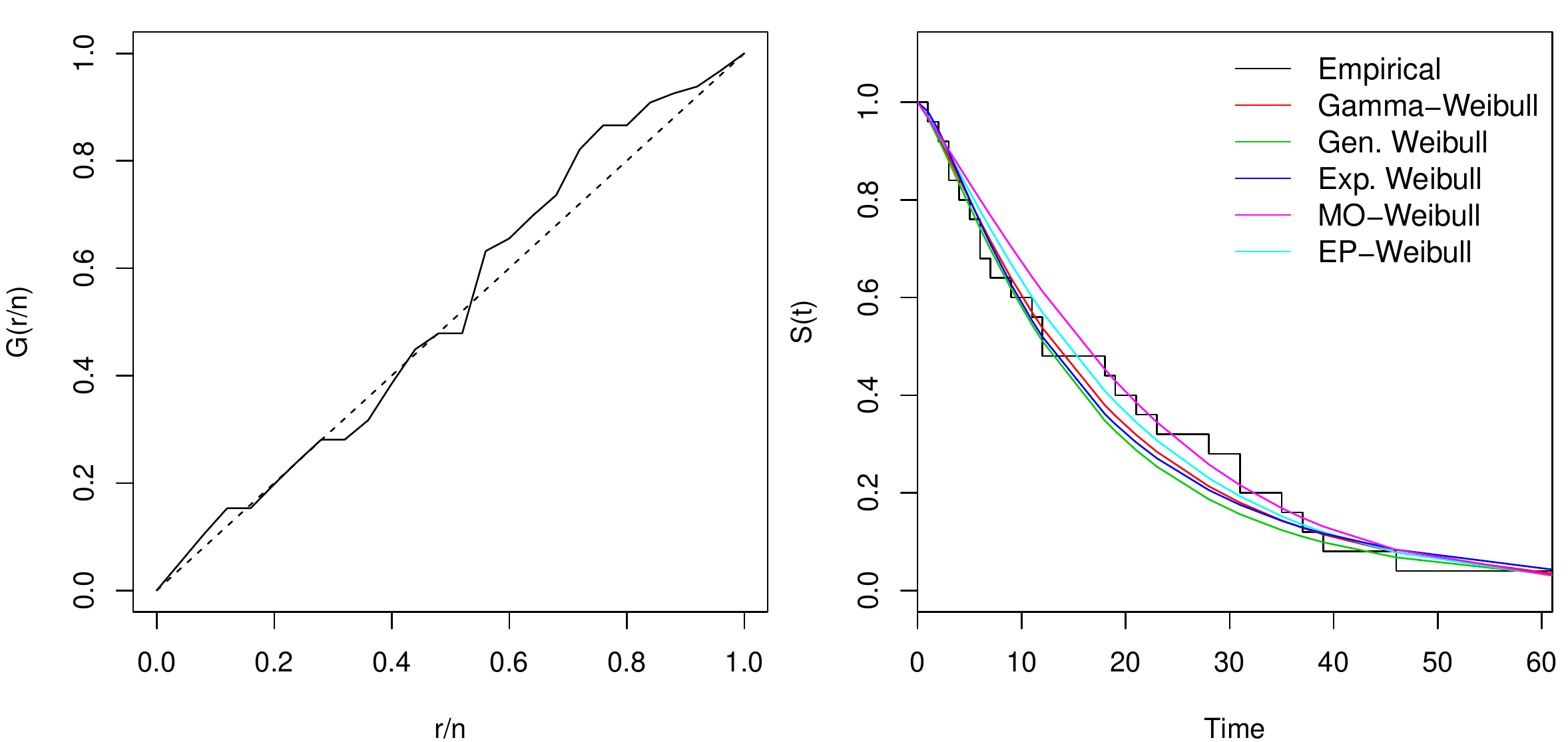}
\caption{Transmission B empirical (left-panel) TTT-plot, and (right-panel) contains the fitted survival superimposed to the empirical survival function and the hazard function adjusted by distribution}\label{graf-transmissor-b}
\end{figure}

Once again, Figure \ref{graf-transmissor-b} presents the TTT-plot, as well as, the survival function fitted by different generalizations of the Weibull distribution, considering the transmission from machine B.

From the TTT-plot we observed that the proposed data may be fitted by all the proposed models since has unimodal hazard rate. Table \ref{tableaic2tB} presents the results of AIC, and AICc in order to discriminate the best fit. 

\begin{table}[ht]
\caption{Results of AIC and AICc criteria and the p-value from the KS test for all fitted distributions considering the Transmission B.}
\centering 
\begin{center}
  \begin{tabular}{ c | c | c | c | c | c }
    \hline
		Criteria & \ Gen. Gamma  \ & \ Gen. Weibull &  \ Exp. Weibull &  \ MO Weibull &  \ EWP\\ \hline
      AIC & 202.220 & 203.201 & 202.833 & 202.368 & 201.997 \\ \hline
			AICc & 197.363 & 198.344 & 197.975 & 197.511 & 203.140 \\ \hline
			KS   & 0.9382 &  0.7657 & 0.8732  & 0.7710 & 0.9622 \\ \hline
  \end{tabular}\label{tableaic2tB}
\end{center}
\end{table}

As shown in Table \ref{tableaic2tB} the EWP distribution has the minimum AIC and AICc. Therefore, we computed its respectively the maximum likelihood estimates and the predictive value for $25\%$. Table \ref{repretransB} presents the MLEs, Standard deviations and $95\%$ credibility intervals for $\alpha$, $\theta$, $\sigma$  and $y*$ related to the GG distribution.

\begin{table}[!h]
	\caption{MLEs, Standard deviations and  $95\%$ credibility intervals for $\alpha$, $\theta$, $\sigma$  and $y*$ related to the GG distribution}
	\centering 
{\small
	{\begin{tabular}{ c | c |  c| c }
			\hline
			$\boldsymbol{\theta}$ & MLE & SD & CI$_{95\%}(\boldsymbol{\theta})$ \\ \hline
			\ \ $\beta$ \ \    & 0.022 &  0.034  &  ( 0.0137; 0.1350)  \\ \hline
			\ \ $\lambda$  \ \   & -0.572 &  0.541  &  (-1.2492; 1.1886)  \\ \hline
			\ \ $\alpha$  \ \   & 1.206 &  0.137  &  ( 0.7579; 1.3705)  \\ \hline
			\ \ $y*$   \ \        & 6.748 &   1.387  &  ( 3.8716; 9.5585)  \\ \hline
		\end{tabular}}\label{repretransB} }
\end{table}

Table \ref{repretransB} results suggest that predictive maintenance should be done in approximately 7 days, considering a point estimation, or given a 95\% credibility interval would be between 4 to 10 days approximated.

\section{Final Remarks}

In this study, we considered different distributions to describe the lifetime of harvest sugarcane machine components. The harvesters stand out for having a large number of corrective stops, given the functionality in such extreme environmental conditions. However, these harvesters does not have an effective preventive maintenance policy which affects its working time schedule. To overcome this problem, we presented a predictive analysis using probability models based on its percentiles aiming to incorporate intelligence into maintenance planning. 

The Weibull distribution is a popular model that can be used to describe a wide range of problems, however, it can not be used to describe data with non-monotone hazard rate. Thus, many generalizations of the Weibull distribution have been proposed to overcome this problem. Since the proposed datasets have non-monotone hazard rate, we considered some flexible generalizations such as the Gamma-Weibull, the generalized Weibull, the Exponentiated Weibull, Marshall-Olkin Weibull and the Extended Poisson Weibull distribution. For the proposed distributions, some mathematical functions were discussed as well as the parameter estimators under the maximum likelihood approach.

The proposed distributions were used to fit the datasets using maximum likelihood estimators. The exponential Weibull presented a superior fit for both machines considering the pricker component, in these cases we concluded a predictive maintenance should be done in approximately 3 days. On the other hand, for the transmission component, the distributions that presented better fit were respectively the Gamma-Weibull distribution and the extended Poisson Weibull for the machine A and B, where a predictive maintenance should be done respectively in 3 and 7 days after the last failure.

Further work should be considered beyond the adjusted models adding to the structure of recurrent event data, implement them, and analyzed their forecast accuracy. This approach should be implemented as an APP, helping the maintenance section in their individualised scheduling distributions.

\section*{Disclosure statement}

No potential conflict of interest was reported by the author(s)

\section*{Acknowledgements}


The research was partially supported by CNPq, FAPESP and CAPES of Brazil.

\bibliographystyle{tfs}

\bibliography{reference}

\end{document}